\documentclass[aps,prl,twocolumn,reprint,superscriptaddress]{revtex4-1}
\usepackage{CJK}
\usepackage[utf8]{inputenc}
\usepackage[english]{babel}
\usepackage{amssymb,amsmath,latexsym,mathrsfs}
\usepackage{feynmp-auto}
\usepackage{graphicx}
\usepackage{makeidx}
\usepackage{subfigure}
\usepackage{epsfig}
\usepackage{multirow}
\usepackage{array}
\usepackage{color}
\usepackage{longtable}
\usepackage{multirow}
\usepackage{bigstrut}
\usepackage{hyperref}
\usepackage{bm}
\usepackage{xcolor}
\usepackage[normalem]{ulem}
\renewcommand{\vec}[1]{\boldsymbol{\mathbf{#1}}}

\newcommand{\sovast}{Sov. Astr.}

\newcommand{\bfchi}{\boldsymbol{\chi}}
\newcommand{\bfK}{\boldsymbol{K}}

\newcommand{\bfD}{\boldsymbol{\Delta}}

\newcommand{\bfDt}{\boldsymbol{\widetilde\Delta}}

\newcommand{\polvec}{%
\begin{pmatrix} Q \\ U \\ V \end{pmatrix}
}
\newcommand{\sqtwo}{\sqrt{2}}

\newcommand{\Vt}{\widetilde{V}}
\newcommand{\ls}{\mathrm{LS}}
\newcommand{\Pp}{\Delta_{P+}}
\newcommand{\Pm}{\Delta_{P-}}
\newcommand{\Ppt}{\widetilde{\Delta}_{P+}}
\newcommand{\Pmt}{\widetilde{\Delta}_{P-}}
\newcommand{\calG}{{\mathcal G}}
\newcommand{\calB}{{\mathcal B}}
\newcommand{\calH}{{\mathcal H}}
\newcommand{\calK}{{\mathcal K}}
\newcommand{\calW}{{\mathcal W}}
\newcommand{\calZ}{{\mathcal Z}}
\newcommand{\wt}[1]{\widetilde{#1}}

\newcommand{\corr}[1]{\langle #1 \rangle}

\newcommand{\threej}[6]{%
\begin{pmatrix} 
#1 & #2 &#3\\
#4 & #5 &#6\\
\end{pmatrix} 
}
\providecommand{\CAMB}{{\tt camb}}
\providecommand{\COSMOMC}{{\tt CosmoMC}}

\providecommand{\CLASS}{{\tt class}}

\providecommand{\GETDIST}{{\tt GetDist}}

\begin{document}

%\title{Through a dark crystal: CMB polarization as a tool to constrain the optical properties of the Universe}
\title{CMB polarization as a tool to constrain the optical properties of the Universe}
\author{Margherita Lembo}
\email{Corresponding author: margherita.lembo@unife.it}
\affiliation{Dipartimento di Fisica e Scienze della Terra, Universit\`a degli Studi di Ferrara, via Giuseppe Saragat 1, I-44122 Ferrara, Italy}
\affiliation{Istituto Nazionale di Fisica Nucleare, Sezione di Ferrara, via Giuseppe Saragat 1, I-44122 Ferrara, Italy}
\author{Massimiliano Lattanzi}
\affiliation{Istituto Nazionale di Fisica Nucleare, Sezione di Ferrara, via Giuseppe Saragat 1, I-44122 Ferrara, Italy}
\author{Luca Pagano}
\affiliation{Dipartimento di Fisica e Scienze della Terra, Universit\`a degli Studi di Ferrara, via Giuseppe Saragat 1, I-44122 Ferrara, Italy}
\affiliation{Istituto Nazionale di Fisica Nucleare, Sezione di Ferrara, via Giuseppe Saragat 1, I-44122 Ferrara, Italy}
\author{Alessandro Gruppuso}
\affiliation{Istituto Nazionale di Astrofisica-Osservatorio di Astrofisica e Scienza dello Spazio, via Gobetti 101, I-40129 Bologna, Italy}
\affiliation{Istituto Nazionale di Fisica Nucleare, Sezione di Bologna, viale Berti Pichat 6/2, I-40127, Bologna, Italy}
\author{Paolo Natoli}
\affiliation{Dipartimento di Fisica e Scienze della Terra, Universit\`a degli Studi di Ferrara, via Giuseppe Saragat 1, I-44122 Ferrara, Italy}
\affiliation{Istituto Nazionale di Fisica Nucleare, Sezione di Ferrara, via Giuseppe Saragat 1, I-44122 
Ferrara, Italy}
\author{Francesco Forastieri}
\affiliation{Dipartimento di Fisica e Scienze della Terra, Universit\`a degli Studi di Ferrara, via Giuseppe Saragat 1, I-44122 Ferrara, Italy}
\affiliation{Istituto Nazionale di Fisica Nucleare, Sezione di Ferrara, via Giuseppe Saragat 1, I-44122 
Ferrara, Italy}

\begin{abstract}
We present a novel formalism to describe the \emph{in vacuo} conversion between polarization states of propagating radiation, also known as generalized Faraday effect (GFE), in a cosmological context.  Thinking of GFE as a potential tracer of new, isotropy- and/or parity-violating physics, we apply our formalism to the cosmic microwave background (CMB) polarized anisotropy power spectra, providing a simple framework to easily compute their observed modifications. In so doing, we re-interpret previously known results, namely the \emph{in vacuo} rotation of the linear polarization plane of CMB photons (or cosmic birefringence) but also point out that GFE could lead to the partial conversion of linear into circular polarization. We notice that GFE can be seen as an effect of light propagating in an anisotropic and/or chiral medium (a ``dark crystal") and recast its parameters as the components of an effective ``cosmic susceptibility tensor". For a wave number-independent susceptibility tensor, this allows us to set an observational bound on a GFE-induced CMB circularly polarized power spectrum, or $VV$, at $C_{\ell}^{VV} < 2 \times 10^{-5} \mu K^2$ (95 \% C.L.), at its peak $\ell\simeq 370$, which is some 3 orders of magnitude better than presently available direct $VV$ measurements. We argue that, unless dramatic technological improvements will arise in direct $V$-modes measurements, cosmic variance-limited linear polarization surveys expected within this decade should provide, as a byproduct, superior bounds on GFE-induced circular polarization of the CMB. 
\end{abstract}

\maketitle

\paragraph{Introduction.} Polarization of the cosmic microwave background (CMB) is the main observational target of next-generation CMB experiments \cite{Ade:2018sbj, litebird2019, Abazajian:2016yjj}. CMB photons are expected to be linearly polarized  by Compton scattering at the epochs of recombination and reionization \cite{Kamionkowski_1997}. In contrast, circular polarization is not expected to be present at the time of last scattering. It can be generated by known physics as CMB photons propagate across the Universe \cite{Kosowsky_1996, Cooray_2003, Giovannini_2009, De:2014qza, Montero-Camacho:2018vgs,Lemarchand:2018lfy, Ejlli:2018ucq}, but only in tiny amounts. However, new physics beyond the standard model of particle physics might be responsible for the generation of a larger amount of circular polarization \cite{Zarei_2010, Kostelecky:2007zz, Alexander:2008fp, Alexander:2019sqb, Alexander_2019, Sadegh_2018, Ejlli:2016avx, Ejlli:2017uli, Inomata_2019, Vahedi_2019, Bartolo_2019,PhysRevD.102.063501}. Observing circular polarization in the CMB, therefore, could provide evidence for new physics.
\\
In this Letter, we introduce a phenomenological framework describing the mixing of CMB polarization states during propagation, including the generation of circular polarization from a pure linearly polarized initial state. We refer to such a mixing as ``generalized Faraday effect'' (GFE) \cite{2013LNP...854.....M, 2005epdm.book.....M}
\footnote{A note on terminology. ``Birefringence'' refers generically to the fact that wave normal modes propagate at different velocities \cite{Carroll:1989vb}. In the cosmological literature, the term ``cosmic birefringence'' usually describes the specific case of different propagation velocity of circular polarization states, leading to the rotation of the linear polarization plane. In crystal optics, the same effect is dubbed ``optical activity''. In the case of a magnetic medium, Faraday rotation refers again to the rotation of the plane of the linear polarization effect, while Faraday conversion denotes the generation of circular from linear polarization. In this Letter, we use generically the terms ``Faraday rotation'' and ``Faraday conversion'' without implying the presence of a magnetic field.
}. 
We use this formalism to derive formulas that allow us to compute the ``observed'' CMB angular power spectra (including the ones involving the circular polarization, $V$) from the ones that would be observed if GFE were absent. The latter can be easily obtained from a Boltzman code like \CAMB\ \cite{Lewis:1999bs} or \CLASS\ \cite{lesgourgues2011cosmic}. We further draw inspiration from the propagation of light in anisotropic or chiral media (e.g., crystals) to describe GFE as the result of light propagating into a medium with an anisotropic and/or parity-violating susceptibility tensor \cite{Gubitosi:2011ue}, and derive constraints on the components of this effective susceptibility tensor from current data. We also discuss the potential of future experiments in this respect.
\paragraph{Theoretical setup.} The transfer equation for polarized radiation in a weakly anisotropic nonabsorbing \footnote{As a concrete example of a model in which one has GFE but not absorption, see e.g. Refs.~\cite{Alexander:2008fp, Bartolo_2019}} medium reads \cite{1968R&QE...11..731S,1969SvA....13..396S,2013LNP...854.....M}
\vspace{-0.2cm}
\begin{equation}\label{eq:radtransfer}
\frac{d}{ds} \polvec = \begin{pmatrix} \epsilon_Q \\ \epsilon_U \\ \epsilon_V \end{pmatrix}  - 
\begin{pmatrix}
0&\rho_V  &-\rho_U \\ 
-\rho_V &0 &\rho_Q \\ 
\rho_U &-\rho_Q  &0
\end{pmatrix} \polvec \, ,
\end{equation}
where $Q,\,U,\,V$ are the Stokes parameters, $s$ is an affine parameter along the photon path, the $\epsilon$'s are the spontaneous emissivities,
and the $\rho$'s describe GFE: $\rho_V$ mixes $Q$ and $U$ polarization and is thus responsible for Faraday rotation,
while $\rho_Q$ and $\rho_U$ mix linear polarization with $V$ and are responsible for Faraday conversion. 
These quantities depend, in principle, on conformal time $\tau$, position $\vec x$ and radiation wave number $\vec p$. In the following we will use the comoving wave number $\vec q = a \vec p$, with $a(\tau)$ being the cosmological scale factor, to describe the wave number dependence. 

CMB linear polarization is sourced by Thomson scattering at the epochs of recombination and reionization sources. From now on we shall assume that the $V$ emissivity is always zero, while the $Q$ and $U$ emissivities are strongly peaked at the time $\tau_\mathrm{rec}$ of hydrogen recombination but vanishing elsewhere, and study Eq.~(\ref{eq:radtransfer}) with $\vec\epsilon = 0$ for $t>t_\mathrm{rec}$ and suitable initial conditions at recombination. Thus we neglect the linear polarization generated at the time of cosmic reionization. We also neglect the effect of gravitational lensing due to matter distribution along the line of sight. We will come back to these approximations later.

When $\vec \epsilon = 0 $, the equation for the (polarization-only) Stokes vector $\mathbf S\equiv \left(Q,\,U,\,V\right)$ can be  recast as
\begin{equation}\label{eq:radtransfer2}
\frac{d\vec S}{ds} =\vec \rho \wedge \vec S \,,
\end{equation}
\vspace{-0.1cm}
where we have introduced the vector in polarization space $\vec\rho\equiv \left( \rho_Q,\, \rho_U , \, \rho_V \right) $, while $\wedge$ represents the usual cross product.\\
In this form, the equation lends itself to a simple interpretation: the total polarization intensity $P= |S|=(Q^2+U^2+V^2)^{1/2}$ is conserved and the vector $\vec S$ precedes with angular velocity $\left|\vec \rho\right|$ around the direction of $\vec \rho$, $ \hat{\vec\rho}$. In other words, if $\hat{\vec \rho}$ is not changing, after some time $t$, the vector $\vec S$ will have been rotated by an angle $2 \alpha \equiv \int \left| \vec \rho \right| dt'$ around $\hat{\vec \rho}$ \footnote{We define the precession angle with a factor 2 so that, in the Faraday rotation limit, $\alpha$ is the angle of rotation of the linear polarization plane.}. 
The direction of $\vec \rho$ defines the polarization of the natural modes of the medium: when $\vec \rho$ is aligned along the $V$ direction, the normal modes are circularly polarized waves; when $\vec \rho$ is orthogonal to the $V$ direction the normal modes are linearly polarized. In general, the normal modes are elliptically polarized states.

Let us briefly comment about the behavior of Eq.~(\ref{eq:radtransfer2}) under rotations around the direction of light propagation -- more pragmatically, under changes in the orientation of the polarimeter. $V$ is a (pseudo)scalar, but $Q$ and $U$ depend on the choice of the reference frame used to measure polarization. When this frame is rotated by an angle $\alpha$, the polarization vector rotates by an angle $2\alpha$ around the $V$--\,axis in polarization space. Then, for the transfer equation to behave covariantly under such transformations, both $\vec\epsilon$  and $\vec\rho$ need to transform in the same way as $\vec S$; in other words, they should be regarded as ``proper'' vectors in polarization space. This is immediately evident by looking at the transfer equation in the form~(\ref{eq:radtransfer2}).

In order to work with quantities with definite spin we introduce the auxiliary polarization vector $\bfD_P\left(\tau,\,\vec x,\, \vec q\right) = \Big( \Delta_{P+}, \Delta_{P-}, V \Big)$, where  $\Delta_{P\pm} =  (Q\pm i U)/\sqtwo$ are the usual spin $\pm 2$ combinations of $Q$ and $U$. 
We similarly define $\rho_\pm \equiv ( \rho_Q \pm i \rho_U)/\sqrt{2}$; from the considerations above, it follows that these should also be spin $\pm 2$ quantities, respectively, while $\rho_V$ is a pseudoscalar.

From Eq.~(\ref{eq:radtransfer2}), we can write down the transfer equation for $\bfD_P$ in a perturbed Friedmann-Robertson-Walker Universe. Expanding the spatial dependence of $\bfD_P$ in Fourier modes and keeping terms up to first order in cosmological perturbations yields
\begin{equation}
\frac{\partial \bfD_P}{\partial \tau} + i k \mu \bfD_P = i \bfK \bfD_P \, ,
\label{eq:tr_master}
\end{equation}
where we have taken into account that the background radiation field is unpolarized. This equation is valid both for scalar and tensor modes. Here, $\vec{k}$ is the
wavevector of the perturbation,  $\mu\equiv\vec{\hat k}\cdot \vec{\hat q}$, and 
we have defined the Hermitian matrix
\vspace{-0.25cm}
\begin{equation}\label{eq:Kmatrix}
\bfK \equiv f(\tau) \begin{pmatrix}
\rho_{0,V}& 0 & -\rho_{0,+} \\
0 & -\rho_{0,V} & \rho_{0,-} \\ 
- \rho_{0,-} & \rho_{0,+} & 0 
\end{pmatrix} \,,
\vspace{-0.25cm}
\end{equation}
where $f(\tau)$ is a generic function of time while the $\rho_{0,X}$'s are time-independent coefficients. 
In deriving Eq.~(\ref{eq:tr_master}), we have also assumed that the $\rho$'s do not depend on $\vec x$. This is equivalent to require that the physics behind the generalized Faraday effect preserves homogeneity \footnote{If the $\bfK$ matrix were to depend on spatial coordinates, the RHS of Eq.~(\ref{eq:tr_master}) would be a convolution (in $k$-space) between $\bfK$ and the Stokes vector. Thus homogeneity has been assumed, explicitly or otherwise by many if not all previous analyses, in which the Fourier modes of the polarization perturbations evolve independently; see for example Refs.~\cite{Alexander:2008fp,Alexander:2019sqb, Bartolo_2019}}.

A formal solution to Eq.~(\ref{eq:tr_master}) with given initial conditions at last scattering $\bfD_P(\tau_\ls) = \bfD_{P,\ls}$ is \footnote{The quantity~(\ref{eq:Delta_sol}) is a solution only if $ \big[ \bfK(t_1) \,, \bfK(t_2) \big] = 0$, which is the case if $\bfK = f(\tau) \bfK_0$. If instead $\bfK$ depends on time in a more general way, the integral at the exponential should be replaced with the Magnus expansion for $\bfK$.}:
\begin{equation}
\vec \Delta_P(\tau) = \exp\left({i \int_{\tau_\ls}^{\tau} \bfK d\tau' }\right)  \bfDt_P (\tau)\,,
\label{eq:Delta_sol}
\end{equation} 
where $\bfDt_P$ solves the initial value problem
\begin{equation}
\frac{\partial \bfDt_P}{\partial \tau} + i k \mu \bfDt_P = 0\, ;\qquad \bfDt_P(\tau_\ls) = \bfDt_\ls \, .
\end{equation}
We get the following relations between the components of $\bfD$ and $\bfDt$ at a given time:
\begin{widetext}
\vspace{-0.5cm} 
\begin{align}\label{eq:master}
&\Pp(\tau) = \Ppt + i\, \sin 2\alpha(\tau) \, \bar \rho^{ -1} (\bar \rho_V \Ppt - \bar \rho_+ \Vt) + \left[1-\cos 2\alpha(\tau)\right]\, \bar \rho^{-2}\left[(\bar \rho_+ \bar \rho_- - \bar \rho^2) \Ppt +  \bar\rho_+^2 \Pmt + \bar\rho_+ \bar\rho_V  \Vt \right] \,; \nonumber \\
&\Pm(\tau) = \Pmt - i\, \sin 2\alpha(\tau) \, \bar\rho^{-1}(\bar\rho_V \Pmt - \bar\rho_- \Vt) + \left[1-\cos2 \alpha(\tau)\right] \, \bar\rho^{-2}\left[\bar\rho_-^2 \Ppt + (\bar\rho_+\bar\rho_- -\bar\rho^2) \Pmt+ \bar\rho_- \bar\rho_V  \Vt \right] \,; \nonumber  \\
&V(\tau) = \Vt -i\, \sin 2\alpha(\tau) \, \bar\rho^{-1}(\bar\rho_- \Ppt - \bar\rho_+ \Pmt) + \left[1-\cos 2\alpha(\tau)\right]\, \bar\rho^{-2}\left[\bar\rho_-\bar\rho_V \Ppt + \bar\rho_+\bar\rho_V \Pmt + (\bar\rho_V^2 -\bar\rho^2) \Vt \right],
\end{align}
\end{widetext}
\vspace{-0.9cm} 
where we are using a bar to denote quantities averaged along the line of sight, i.e., $\bar\rho_X \equiv (\tau-\tau_\ls)^{-1}\rho_{0,X}\,\int_{\tau_\ls}^\tau f(\tau') \, d\tau'$, $\bar \rho^2 \equiv \bar \rho^2_Q + \bar \rho^2_U + \bar \rho^2_V = 2 \bar \rho_+ \bar \rho_- + \bar \rho^2_V$ and $2\alpha (\tau) = \bar \rho (\tau - \tau_{LS})$. 
These equations express the polarization perturbations after the mixing (the ``un-tilded'' quantities appearing in the left-hand side) 
in terms of those that would be realized in the sky if such a mixing were absent (the tilded quantities appearing on the right-hand side). 
They can be seen as a generalization of the equations for anisotropic cosmic birefringence, that is sourced by $Q-U$ mixing, to the
case of a $Q-U-V$ mixing.

In order to characterize the statistics of the CMB perturbations, we need to calculate angular power spectra. We thus expand in spherical harmonics both sides of Eqs.~\eqref{eq:master}. 
We take $\widetilde V = 0$, coherently with the standard expectation of the vanishing primordial $V$ mode.
While $V,\,\widetilde V$,  and $\bar \rho_V$ are scalar quantities and can be naturally expanded in spin-0 spherical harmonics, $\Delta_{P\pm}$, $\widetilde{\Delta}_{P\pm}$, and $\bar\rho_\pm$ should be expanded in spin-weighted $s=\pm 2$ harmonics \cite{Zaldarriaga_1997}. We denote the expansion coefficients of $\Delta_{P\pm}$, $V$ and $\widetilde{\Delta}_{P\pm}$ as $a_{\pm /V, \ell m}$ and $\widetilde a_{\pm, \ell m}$, while the expansion coefficients  of $\bar \rho_{\pm/ V}(\tau-\tau_{LS})$
as $b_{\pm 2/V, \ell m}$ respectively.   \\
Projecting both sides of Eqs.~(\ref{eq:master}) over the appropriate spherical harmonics and keeping only terms up to second order in $\alpha$,  we obtain:
\begin{align} \label{eq:exp_coeff}
a_{E,L} &= \widetilde a_{E,L} + \left(\calG^{(1)}_{L_1 L} - \calH^{(1)}_{L_1 L} + \calH^{(3)}_{L_1 L} \right)  \widetilde a_{E,L_1} + \nonumber \\
&- \left(\calG^{(2)}_{L_1 L} + \calH^{(2)}_{L_1 L} - \calH^{(4)}_{L_1 L}\right) \widetilde a_{B,L_1} \,; \nonumber \\
a_{B,L} &= \widetilde a_{B,L} + \left(\calG^{(1)}_{L_1 L} - \calH^{(1)}_{L_1 L} - \calH^{(3)}_{L_1 L} \right)  \widetilde a_{B,L_1} + \nonumber \\
&+ \left(\calG^{(2)}_{L_1 L} + \calH^{(2)}_{L_1 L} + \calH^{(4)}_{L_1 L}\right) \widetilde a_{E,L_1} \,; \nonumber \\
a_{V,L} &= \left(  \calG^{(3)}_{L_1 L} -  \calH^{(5)}_{L_1 L} \right) \widetilde a_{E,L_1} -  \left( \calG^{(4)}_{L_1 L} + \calH^{(6)}_{L_1 L} \right) \widetilde a_{B,L_1} \,,
\end{align}
where we have introduced $a_{E,L} = - (a_{2,L} + a_{-2,L})/2$ and  $a_{B,L} = i (a_{2,L} - a_{-2,L})/2$, 
$L$ stands for $(\ell,m)$, and summation over repeated indices is understood.
The $\calG$ and $\calH$ kernels contain geometrical factors (products of Wigner-$3j$ symbols) and the $b$ expansion coefficients at the first and the second orders respectively, but do not depend on the other cosmological parameters. Their explicit form is given in the Supplemental Material.\\

We can use Eqs.~(\ref{eq:exp_coeff}) to build correlators $C^{XY}_{LL'}\equiv \corr{a^{\phantom{*}}_{X,L} a^*_{Y,L'}}$.
We focus on the diagonal components ($L=L'$), but in principle there is potentially valuable information also in the off-diagonal terms. Defining $C^{XY}_\ell \equiv \left( 2\ell+1 \right)^{-1} \sum_{m=-\ell}^{+\ell} C^{XY}_{LL}$, we get the following expressions for the GFE-modified angular power spectra:
\begin{align}\label{eq:spectra}
C^{TE}_\ell &= \left(1- \frac{\calZ}{2} \right) \wt{C}^{TE}_{\ell}  \,;\quad C^{TB}_\ell = \frac{b_{V,00}}{\sqrt{4\pi}}\wt{C}^{TE}_{\ell}  \,; \nonumber\\[0.2cm] 
C^{EE}_\ell &= \left(1-\calZ\right) \wt{C}^{EE}_{\ell} + \calK^{11}_{\ell_1\ell}  \wt{C}^{EE}_{\ell_1} + \calK^{22}_{\ell_1\ell}  \wt{C}^{BB}_{\ell_1}  \,;\nonumber \\[0.2cm]
C^{BB}_\ell &= \left(1-\calZ\right) \wt{C}^{BB}_{\ell} + \calK^{11}_{\ell_1\ell}  \wt{C}^{BB}_{\ell_1} + \calK^{22}_{\ell_1\ell}  \wt{C}^{EE}_{\ell_1} \,; \nonumber \\[0.2cm]
C^{EB}_\ell &= \frac{b_{V,00}}{\sqrt{4\pi}}\left(\wt{C}^{EE}_{\ell}-\wt{C}^{BB}_{\ell}\right)  \,; \nonumber\\[0.2cm]
C^{EV}_\ell &= \calK^{13}_{\ell_1\ell}\wt{C}^{EE}_{\ell_1}+\calK^{24}_{\ell_1\ell}\wt{C}^{BB}_{\ell_1} \,; \nonumber \\[0.2cm]
C^{BV}_\ell &= \calK^{23}_{\ell_1\ell}\wt{C}^{EE}_{\ell_1}-\calK^{14}_{\ell_1\ell}\wt{C}^{BB}_{\ell_1} \,; \nonumber \\[0.2cm]
C^{VV}_\ell &= \calK^{33}_{\ell_1\ell}\wt{C}^{EE}_{\ell_1}+\calK^{44}_{\ell_1\ell}\wt{C}^{BB}_{\ell_1} \,,
\end{align}
where we have defined the $\calK$ kernels as
$\calK^{a\,b}_{\ell_1\ell}= \left(2\ell+1\right)^{-1} \sum_{m_1, m} \calG^{(a)}_{L_1 L}\calG^{(b)*}_{L_1 L} \,$, and $4\pi\calZ=\sum_{\ell m} \left(\left| b_{V, \ell m}\right|^2 + \left| b_{2, \ell m}\right|^2 \right)$ \footnote{Note that if the $\rho$'s have to be interpreted as stochastic quantities, $\calZ$ gives their variance.}.
In deriving these equations, we have assumed the absence of primordial $TB$ and $EB$ correlations.\\
The spectra that do not appear in Eqs.~(\ref{eq:spectra}) are equal to their ``unrotated'' counterparts to second order in $\alpha$. 
We note that the mixing among the polarization components possibly leads to a nonzero $VV$ power spectrum as well as to parity-violating power spectra, such as $EB$, $TB$ and $EV$.

Equations (\ref{eq:spectra}) encode, in a very general way, the modifications due to GFE, linking the modified power spectra to the power spectra that we would have in absence of this effect: they follow quite generally from Eq.~(\ref{eq:radtransfer}). 

It is interesting to consider some limiting cases of Eqs.~(\ref{eq:spectra}). When only $\rho_V\neq0$, the Stokes vector rotates around the $V-$direction and only $Q$ and $U$ mix; this is the so-called ``cosmic birefringence'', widely studied in the literature. It is immediate to convince oneself that in this case $\alpha$ is the birefringence angle, i.e., the angle of rotation of the plane of linear polarization, and to recover, from Eqs.~(\ref{eq:spectra}), the equations for both isotropic and anisotropic birefringence at second order in $\alpha$ ~\cite{Gruppuso_2016, Li:2008tma, Zhao:2015mqa}.\\ 
Similarly, when $\rho_Q \neq0$ and/or $\rho_U \neq0 $, we recover Faraday conversion, i.e., circular polarization is generated by \emph{conversion} of the primordial linear polarization as CMB photons propagate through a birefringent medium along the line of sight \cite{Montero-Camacho:2018vgs, Kamionkowski_2018}. 

Mixing of polarization components can arise through several mechanisms, involving either known physics or more exotic models. Given such a mechanism, the $\rho's$ can be computed and specific predictions for the observed power spectra can be obtained.
In order to see our formalism at work, we adopt an agnostic point of view and relate the GFE parameters to the optical properties of the medium traversed by CMB photons.

The optical properties of a medium depend on the three-dimensional (dielectric) susceptibility tensor $\bfchi$.
The more general form of the dielectric tensor of a nondispersive medium is \footnote{The reference frame in which the susceptibility tensor takes this form is the one aligned with the principal axes of the crystal, i.e., the eigenvectors of $\Re({\chi})$.}
\begin{equation}
\vec{\chi} = 
\begin{pmatrix}
\chi_{xx} & i \,\chi_{xy} & -i\,\chi_{xz} \\
- i\, \chi_{xy} & \chi_{yy} & i\, \chi_{yz} \\
i\, \chi_{xz} & -i\, \chi_{yz} & \chi_{zz} 
\end{pmatrix} \,,
\label{eq:chi}
\end{equation}
where the $\chi_{i j}$ are all real, so that $\vec{\chi}$ is Hermitian. We assume that the medium is homogeneous, 
therefore $\vec{\chi}$ does not depend on position; it might however depend on the radiation wave number. The diagonal (off-diagonal) elements are responsible for different linear (circular) polarization states propagating with different velocities, and as such they violate isotropy (parity).

To the purpose of making a connection between the three-dimensional susceptibility tensor and the $\rho$'s, we compare Eq.~(\ref{eq:radtransfer}) with the radiative transfer equation written in terms of the susceptibility tensor \cite{1968R&QE...11..731S,1969SvA....13..396S,2005epdm.book.....M}:
\begin{equation}
\left(\frac{\partial}{\partial t} + \vec{\hat p} \cdot\vec{\nabla}\right){\mathcal I}_{ab} = {\mathcal E}_{ab} + i (2\pi \nu)\left( \chi^{(2)}_{ac} {\mathcal I}_{cb} - {\mathcal I}_{ac} (\chi^{(2)\dag})_{cb}\right) \, ,
\label{eq:sazo}
\end{equation}
where ${\mathcal I}_{ab}$ is the polarization tensor, ${\mathcal E}_{ab}$ is the tensor of spontaneous emission intensity per unit volume, $\nu$ is the radiation frequency, and $\chi^{(2)}_{ab}$ is the susceptibility tensor in the plane perpendicular to the direction of light propagation. We then get, for the mixing coefficients in the direction $(\theta,\,\phi)$:
\begin{align}
& \rho_Q =2\pi\nu_0 \Big[(\chi_{xx} c_\theta^2 -\chi_{yy}) c_\phi^2 + (\chi_{yy} c_\theta^2 -\chi_{xx}) s_\phi^2+
\chi_{zz} s_\theta^2  \Big]; \nonumber \\
&\rho_U= 4\pi\nu_0 (\chi_{yy}-\chi_{xx}) c_\theta s_\phi c_\phi; \nonumber \\
&\rho_V = 4\pi\nu_0\left( \chi_{xy} c_\theta + \chi_{yz} s_\theta c_\phi + \chi_{xz} s_\theta s_\phi\right) \,,
\label{eq:rho_of_chi}
\end{align}
where $\nu_0=a\nu$, $s_X \equiv \sin X$ and $c_X\equiv \cos X$, and in general the $\chi_{ij}$'s will themselves depend on $(\theta,\,\phi)$.
Given the above discussion on the connection between the elements of $\vec\chi$ and the symmetries of the medium, this relation makes clear that in an anisotropic medium $\rho_Q\ne0$ and/or $\rho_U\ne0$, while parity violation implies $\rho_V\ne 0$.

A mechanism that alters the propagation of photons across cosmological distances can be recast in terms of an effective dielectric tensor, for example by looking at how the wave equation is modified. Then Eqs.~(\ref{eq:rho_of_chi}) and (\ref{eq:spectra}) can be readily used to obtain predictions for the observed CMB angular power spectra.
We explicit this procedure for a dielectric tensor that does not depend on the radiation wave number \footnote{This amounts to the assumption that the (unknown) natural length of the medium, which is related to the physics underlying GFE, is much smaller than the CMB wavelengths.}. In this case, the modified spectra in Eqs.~(\ref{eq:spectra}) read
\vspace{-0.25cm}
\begin{align}
C^{TE}_\ell & = \wt{C}^{TE}_{\ell} - \frac{1}{2} \left( \frac{\beta^2_V + \beta^2_E}{4 \pi} \right) \wt{C}^{TE}_{\ell}\,;\nonumber\\
C^{EE}_\ell & = \wt{C}^{EE}_{\ell}  - \left( \frac{\beta^2_V + \beta^2_E}{4 \pi} \right) \wt{C}^{EE}_{\ell} + \nonumber \\ 
&+  \frac{\beta^2_V}{4 \pi} \Big[ \calW^{(1)}_\ell\, \wt{C}^{EE}_{\ell} +  \calW^{(1)}_{\ell+1}\, \wt{C}^{BB}_{\ell+1} + \calW^{(1)}_{\ell-1}\, \wt{C}^{BB}_{\ell-1}  \Big]\,;  \nonumber\\
C^{BB}_\ell & = \wt{C}^{BB}_{\ell} -  \left( \frac{\beta^2_V + \beta^2_E}{4 \pi} \right) \wt{C}^{BB}_{\ell} + \nonumber \\
&+   \frac{\beta^2_V}{4 \pi} \Big[ \calW^{(1)}_\ell\, \wt{C}^{BB}_{\ell} +  \calW^{(1)}_{\ell+1}\, \wt{C}^{EE}_{\ell+1} + \calW^{(1)}_{\ell-1}\, \wt{C}^{EE}_{\ell-1}  \Big] \,; \nonumber \\
C^{VV}_\ell & =  \frac{\beta^2_E}{\pi} \,\Big[  \calW^{(2)}_{\ell+2}\, \wt{C}^{BB}_{\ell+2} + \calW^{(2)}_{\ell+1}\, \wt{C}^{EE}_{\ell+1} \nonumber 
 + \calW^{(2)}_\ell \, \wt{C}^{BB}_{\ell} + \\ 
 & + \calW^{(2)}_{\ell-1}\, \wt{C}^{EE}_{\ell-1} + \calW^{(2)}_{\ell-2}\, \wt{C}^{BB}_{\ell-2} \Big] \,,
\label{eq:spectra2}
\end{align}
where $\calW^{(1)}_\ell$ and  $\calW^{(2)}_\ell$ are combinations of Wigner $3j$-symbols, and 
\begin{align}
&\beta^2_V \equiv  \, \sum_{m'} \left|b_{V,1m'}\right|^2  
= \frac{16\pi}{3} \left(\xi_{xy}^{\,2}+ \xi_{yz}^{\,2} + \xi_{xz}^{\,2} \right) ;\nonumber \\
&\beta^2_E \equiv  \, \sum_{m'} \left|\frac{1}{2}(b_{2,\ell m} + b_{-2,\ell m}) \right|^2  =   \nonumber \\
&=\frac{8\pi}{15} \left[\left(\xi_{xx}-\xi_{yy}\right)^{2}+ \left(\xi_{yy}-\xi_{zz}\right)^{2}
+\left(\xi_{zz}-\xi_{xx}\right)^{2}\right] \,,
\label{eq:params}
\end{align}
having defined
$
\xi_{i j} = \int_{\tau_\ls}^{\tau_0} 2\pi \nu_0 \chi_{i j} d\tau \, .
$
Note that $\beta^2_V$ and $\beta^2_E$ probe independent combinations of the $\chi_{i j}$'s, since they only depend on the \emph{off}- and \emph{on}-diagonal components, respectively. 

\begin{figure}
  \includegraphics[width=0.95\columnwidth]{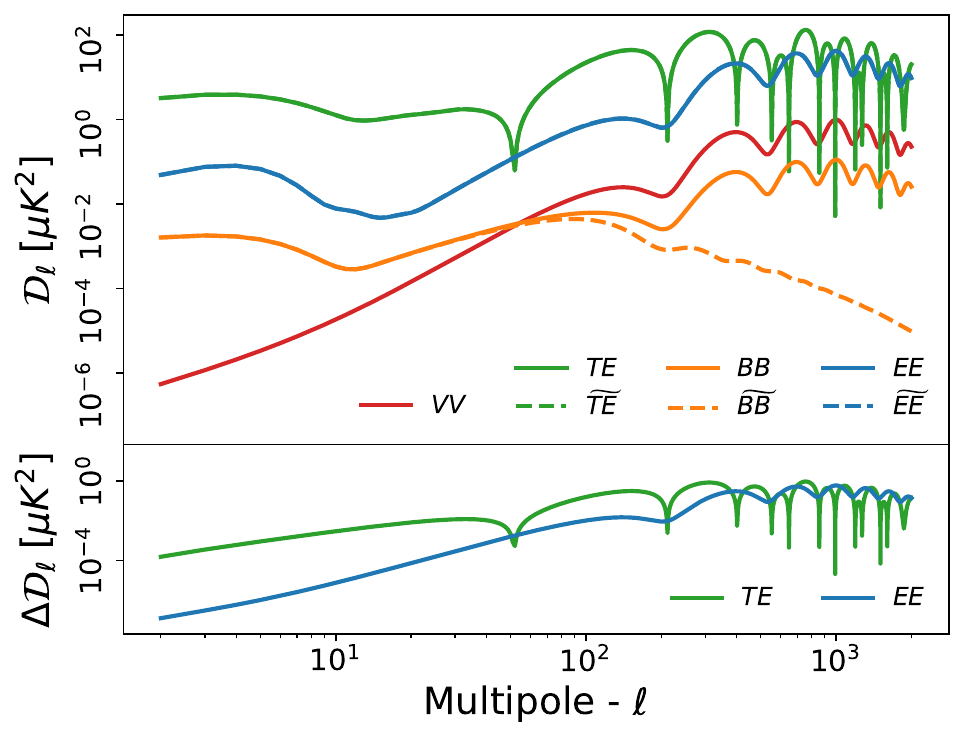}
\caption{\emph{Top.} ``Rotated" (solid line) and ``unrotaded" angular power spectra (dashed line). For the rotated ones we have used $\beta^2_V=0.03$ and $\beta^2_E=0.14$. In both case the other cosmological parameters are the best-fit values of Planck 2018 and $r_{0.05} = 0.07$. \emph{Bottom.} Absolute differences between ``rotated" and ``unrotated" for $EE$ and $TE$. \label{fig:mod_spectra}}
\end{figure}

The effects of these modifications are made clear in Fig.~\ref{fig:mod_spectra}, where we show polarization power spectra computed assuming the best-fit Planck 2018 cosmology \cite{Aghanim:2018eyx}, and the corresponding modified spectra for $\beta_V^2 = 0.03$ and $\beta_E^2 = 0.14$ \footnote{The code used to compute these GFE-induced power spectra is made publicly available on github: \url{https://github.com/mlembo00/circular-polarization.git}}. A nonvanishing $V$-modes power spectrum is generated, whose shape mostly follows that of the $E$-modes spectrum, as expected from Eqs.~(\ref{eq:spectra2}). The $B$-modes power spectrum is dramatically affected even for relatively small values $\beta_V^2$, because power leaks from the much larger $E$-modes. 
Note that in producing the curves in Fig.~\ref{fig:mod_spectra} we only rotate the polarization produced at last scattering. 

\paragraph{Observational constraints.} Observations of CMB polarization can be used to constrain the values of $\beta^2_V$ and $\beta^2_E$, in the framework of simple extensions of the $\Lambda$CDM model. 
We use observations of temperature and linear polarization anisotropies from the Planck legacy release \cite{Aghanim:2019ame}  and BICEP2/Keck 2015 \cite{Ade:2018gkx} to derive bounds on 
both $\beta^2_V$ and $\beta^2_E$ in the presence of primordial tensor modes, parameterized by the tensor-to-scalar ratio $r$. Using the Monte Carlo engine \COSMOMC\ \cite{Lewis:2002ah}, we find $\beta^2_V < 0.030$, $\beta^2_E < 0.14$ and $r_{0.05} < 0.055$ at 95\% CL \footnote{The bound on $r$ is slightly stronger that the corresponding value for $\beta^2_V=0$ due to degeneracy between the two parameters.} In deriving these constraints, we have separated the effect of lensing and GFE. In principle, these two effects might be acting simultaneously and should be treated accordingly. However, in our analysis, we rotate the unlensed CMB power spectra, using Eqs.~\ref{eq:spectra2}, and then we add the lensing contribution as computed by \CAMB. For the noise level of current CMB data, we have checked that rotating instead the lensed spectra leads to consistent results. We thus argue that our treatment of lensing is accurate enough for our purposes. This still holds for the LiteBIRD satellite \cite{litebird2019} and Simons Observatory \cite{Ade:2018sbj}. Regarding fourth-generation ground-based CMB experiments (e.g., CMB-S4) \cite{Abazajian:2019eic}, the interplay between lensing and GFE should be modeled in more detail.

Circular polarization data are also sensitive to $\beta^2_E$, see Eq.~(\ref{eq:spectra2}).
Using the $V$-modes CMB polarization 
data from the CLASS telescope \cite{Essinger-Hileman:2014pja,Class2019}, we find $\beta^2_E < 38$ (95\% C.L.), assuming the Planck 2018 best-fit 
$EE$ and $BB$ spectra. This constrain violates our assumption that $\rho_{Q,U} \ll 1$, nevertheless it indicates that current $V$-mode data allow a large mixing of linear and circular polarization. We have shown that such a large mixing is, however, excluded by current observations of $E$ and $B$ polarization.

These constraints on the $\beta$'s can be recast in terms of the $\chi_{ij}$. Taking $\nu_0\simeq150$~GHz as the frequency of the CMB photons today, 
$\xi_{i j}  \simeq 1.5 \times 10^{30}\,  \chi_{i j}$ if the $\chi$'s do not depend on time. The constraint on $\beta^2_V$ implies a bound $\chi_{i j} \leq 2.7 \times 10^{-32}$ for the largest off-diagonal element,
while the one on $\beta^2_E$ implies $\chi_{i i} - \chi_{j j} \leq 1.3 \times 10^{-31}$ for the largest difference between diagonal elements
\footnote{In the case of a time-dependent susceptibility tensor, the same constraints apply to the time-averages $\overline\chi_{ij} \equiv
(\tau_0-\tau_\ls)^{-1}\int_{\tau_\ls}^{\tau_0} \chi_{ij}(\tau') d\tau'$. }.

\begin{figure}
  \includegraphics[width=0.95\columnwidth]{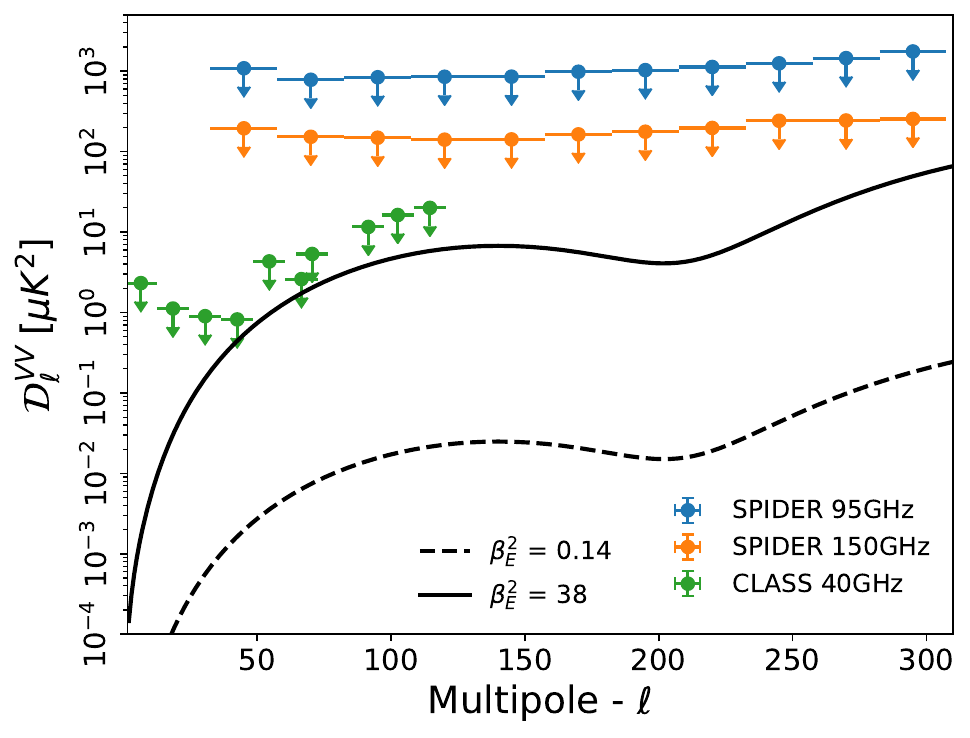}
\caption{$V$-modes power spectra as predicted by CLASS (solid line) and Planck/BICEP2/Keck (dashed line), compared with SPIDER\cite{Nagy:2017csq} and CLASS\cite{Class2019} data.\label{fig:vmodesdata}} 
\end{figure}

\paragraph{Conclusions.} In this Letter, we have derived a transparent, and convenient to use, set of expressions for the observed angular power spectra of CMB polarization, including circular polarization, in the presence of generalized Faraday effect, i.e., the precession of the Stokes polarization vector in $(Q,\,U,\,V)$ space. Equations~(\ref{eq:spectra}) are valid to second order in the small rotation angle of the Stokes vector. To our knowledge, this is the first time that such expressions appear in the literature. We have also proposed a phenomenological framework in which the Universe is regarded as an homogeneous but possibly anisotropic and/or chiral medium for what concerns the propagation of light. The optical properties of the medium are encoded in its susceptibility tensor; models predicting GFE can in principle be recast in such terms. 
We have linked the expressions for GFE to the components of the susceptibility tensor allowing for an easy way to derive predictions for this class of models. Finally, we have derived constraints for a simple benchmark model with a wave number-independent susceptibility tensor. Also in this case, this is the first time that such limits appear in the literature. In Fig.~\ref{fig:vmodesdata} we show the current data on the CMB $VV$ power spectrum, together with theoretical power spectra for $\beta_E^2  = 38$, corresponding to the 95\% upper limit allowed by current circular polarization data, and $\beta^2_E=0.14$, corresponding to the 95\% upper limit allowed by current temperature and linear polarization data.   

Forthcoming experiments, such as CLASS \cite{Class2019}, will likely improve the sensitivity on $V$ mode. However, an improvement in sensitivity by a factor $\sim 10^3$ (at the level of spectra) would be necessary to bring constraints from circular polarization observations at the same level as the current bounds from linear polarization, see Fig.~\ref{fig:vmodesdata}.
Moreover, future linear polarization measurements from LiteBIRD satellite will further improve the constraint on $\beta^2_E$ by roughly a factor 3, down to $\beta^2_E<0.05 $, while a cosmic variance-limited experiment could potentially reach $\beta^2_E < 0.01$.
We thus argue that linear polarization measurements from forthcoming experiments will likely yield stronger constraints on GFE than direct observations of circular polarization, at least in the case of a wavenumber-independent effective susceptibility tensor.

\begin{acknowledgments}
\paragraph{Acknowledgments.}

We acknowledge the use of \CAMB\ \cite{Lewis:2002ah} and \GETDIST\ \cite{Lewis:2019xzd}  software packages, and the use of computing facilities provided by the INFN theory group (I.S. InDark) at CINECA. 
We are grateful to G. Fabbian for useful discussion, to G. Iacobellis for help in an early stage of this work, and to J. Gudmundsson for help with the SPIDER data. We thank the anonymous referees for their insightful comments, which helped further refine our analysis.
We also acknowledge the financial support from the COSMOS network through the ASI (Italian Space Agency) Grants No. 2016-24-H.0, No. 2016-24-H.1-2018, and No. 2019-9-HH.0.
\end{acknowledgments}

\bibliography{bibliography}

\clearpage
\newpage
\onecolumngrid
\setcounter{equation}{0}
\appendix*
\section{SUPPLEMENTAL MATERIAL}
The modifications due to GFE are fully encoded in the following equations, 
\begin{subequations}\label{eq:exp_coeff}
\begin{align} 
a_{E,L} &= \widetilde a_{E,L} + \left(\calG^{(1)}_{L_1 L} - \calH^{(1)}_{L_1 L} + \calH^{(3)}_{L_1 L} \right)  \widetilde a_{E,L_1} - \left(\calG^{(2)}_{L_1 L} + \calH^{(2)}_{L_1 L} - \calH^{(4)}_{L_1 L}\right) \widetilde a_{B,L_1} \,;  \\
a_{B,L} &= \widetilde a_{B,L} + \left(\calG^{(1)}_{L_1 L} - \calH^{(1)}_{L_1 L} - \calH^{(3)}_{L_1 L} \right)  \widetilde a_{B,L_1} + \left(\calG^{(2)}_{L_1 L} + \calH^{(2)}_{L_1 L} + \calH^{(4)}_{L_1 L}\right) \widetilde a_{E,L_1} \,;  \\
a_{V,L} &= \left(  \calG^{(3)}_{L_1 L} -  \calH^{(5)}_{L_1 L} \right) \widetilde a_{E,L_1} -  \left( \calG^{(4)}_{L_1 L} + \calH^{(6)}_{L_1 L} \right) \widetilde a_{B,L_1} \,,
\end{align}
\end{subequations}
where we adopted the usual definition, $a_{E,L} = - (a_{2,L} + a_{-2,L})/2$ and  $a_{B,L} = i (a_{2,L} - a_{-2,L})/2$, and where $L$ stands for $(\ell,m)$. From now on, summation over repeated indices is understood.
\\
The expressions of the $\calG$ and $\calH$ kernels are

\begin{subequations}
\begin{align}
%G
&\calG^{(1)}_{L_1L} = (-1)^m  \,i\sum_{\substack{\ell_2 \\ \ell+\ell_1+\ell_2\,\mathrm{odd} \\ m_2 = m- m_1}} G_{L_1 L_2 L}\, b_{V,L_2};  \\[0.25cm]
&\calG^{(2)}_{L_1L} = (-1)^m \sum_{\substack{\ell_2 \\ \ell+\ell_1+\ell_2\,\mathrm{even} \\ m_2 = m- m_1}} G_{L_1 L_2 L}\, b_{V,L_2};  \\[0.25cm]
&\calG^{(3)}_{L_1L}=(-1)^m \, i\sum_{\substack{\ell_2  \\ m_2 = m- m_1}}
G'_{L_1 L_2 L} \bigg[ \big(b_{-2,L_2} - (-1)^{\ell+\ell_1+\ell_2}\, b_{2,L_2}\big)\bigg]; \\[0.25cm]
&\calG^{(4)}_{L_1L}=(-1)^m \sum_{\substack{\ell_2  \\ m_2 = m- m_1}}
G'_{L_1 L_2 L} \bigg[ \big(b_{-2,L_2} + (-1)^{\ell+\ell_1+\ell_2}\, b_{2,L_2}\big)\bigg] ;  
\end{align}
\begin{align}
%H
&\calH^{(1)}_{L_1L} = \frac{(-1)^m}{2} \sum_{\substack{\ell_2 \\ \ell+\ell_1+\ell_2\,\mathrm{even} \\ m_2 = m- m_1}} G_{L_1 L_2 L}\, \left( \calB_{V,L_2} + \calB_{\pm 2,L_2}\right);  \\[0.25cm]
&\calH^{(2)}_{L_1L} = \frac{(-1)^m}{2} \,i\sum_{\substack{\ell_2 \\ \ell+\ell_1+\ell_2\,\mathrm{odd} \\ m_2 = m- m_1}} G_{L_1 L_2 L}\, \left( \calB_{V,L_2} + \calB_{\pm 2,L_2}\right);  \\[0.25cm]
&\calH^{(3)}_{L_1L}=\frac{(-1)^m}{4} \sum_{\substack{\ell_2  \\ m_2 = m- m_1}}
G''_{L1 L_2 L} \bigg[ \big(\calB_{-4,L_2} + (-1)^{\ell+\ell_1+\ell_2}\, \calB_{4,L_2}\big)\bigg];  \\[0.25cm]
&\calH^{(4)}_{L_1L}=\frac{(-1)^m}{4}\,i \sum_{\substack{\ell_2  \\ m_2 = m- m_1}}
G''_{L_1 L_2 L} \bigg[ \big(\calB_{-4,L_2} - (-1)^{\ell+\ell_1+\ell_2}\, \calB_{4,L_2}\big)\bigg];
\end{align}
\begin{align}
&\calH^{(5)}_{L_1L}=\frac{(-1)^m}{2} \sum_{\substack{\ell_2  \\ m_2 = m- m_1}}
G'_{L1 L_2 L} \bigg[ \big(\calB_{-2V,L_2} + (-1)^{\ell+\ell_1+\ell_2}\, \calB_{2V,L_2}\big)\bigg];  \\[0.25cm]
&\calH^{(6)}_{L_1L}=\frac{(-1)^m}{2}\,i \sum_{\substack{\ell_2  \\ m_2 = m- m_1}}
G'_{L_1 L_2 L} \bigg[ \big(\calB_{-2V,L_2} - (-1)^{\ell+\ell_1+\ell_2}\, \calB_{2V,L_2}\big)\bigg],
\end{align}
\end{subequations}
where
\begin{subequations}
\begin{align}
G_{L_1 L_2 L} \equiv F_{\ell_1 \ell_2 \ell}\,
\threej{\ell_1}{\ell_2}{\ell}{m_1}{m_2}{-m}
\threej{\ell_1}{\ell_2}{\ell}{-2}{0}{2} ; \\[0.5cm] 
G'_{L_1 L_2 L} \equiv F_{\ell_1 \ell_2 \ell}\,
\threej{\ell_1}{\ell_2}{\ell}{m_1}{m_2}{-m}
\threej{\ell_1}{\ell_2}{\ell}{-2}{2}{0} ; \\[0.5cm]
G'_{L_1 L_2 L} \equiv F_{\ell_1 \ell_2 \ell}\,
\threej{\ell_1}{\ell_2}{\ell}{m_1}{m_2}{-m}
\threej{\ell_1}{\ell_2}{\ell}{-2}{2}{0} ; \\[0.5cm]
G''_{L_1 L_2 L} \equiv F_{\ell_1 \ell_2 \ell}\,
\threej{\ell_1}{\ell_2}{\ell}{m_1}{m_2}{-m}
\threej{\ell_1}{\ell_2}{\ell}{2}{-4}{2} .
\end{align}
\end{subequations}
with $F_{\ell_1 \ell_2 \ell} = \left[(2\ell_1+1)(2\ell_2+1)(2\ell+1)/4\pi\right]^{1/2}$.

\end{document}